\documentclass[prb,aps,twocolumn,superscriptaddress]{revtex4-2}
\usepackage{amsmath}
\usepackage{color}
\usepackage{bbm}
\usepackage{amssymb}
\usepackage{epsfig}
\usepackage{multirow}
\usepackage{amsbsy}
\usepackage{array}
\usepackage{diagbox}
\usepackage{bm}
\usepackage{extarrows}
\usepackage{graphicx}
\usepackage{subfigure}
\usepackage{appendix}
\usepackage{txfonts}
\usepackage{lipsum}
\usepackage{bbding}
\usepackage{pifont}
\usepackage{siunitx}
\usepackage{makecell}
\usepackage{gensymb}
\usepackage[version=4]{mhchem}
\graphicspath{{Figures/}}
\allowdisplaybreaks[4]
\usepackage[colorlinks=true,linkcolor=blue,citecolor=blue,urlcolor=blue,bookmarks=false]{hyperref}

\makeatletter

\newif\if@equalcontrib

\newcommand{\equalcontribtext}
{These authors contributed equally to this work.}

\DeclareRobustCommand{\equalcontrib}{%
  \@equalcontribtrue
}

\def\doauthor#1#2#3{%
  \@equalcontribfalse
  \ignorespaces#1\unskip\@listcomma
  \begingroup
    #3%
  \@if@empty{#2}
    {\endgroup{}{}}
    {\endgroup{\comma@space}{}%
     \frontmatter@footnote{#2}}%
  \if@equalcontrib
    \comma@space
    \frontmatter@footnote{\equalcontribtext}%
  \fi
  \space\@listand
}

\makeatother

\begin{document}

\title{Annular Majorana mode in a superconducting topological insulator}

\author{Shengshan Qin\equalcontrib}
\email{qinshengshan@bit.edu.cn}
\affiliation{School of Physics, Beijing Institute of Technology, Beijing 100081, China}

\author{Chi Wu\equalcontrib}
\affiliation{Institute of Theoretical Physics, Chinese Academy of Sciences, Beijing 100190, China}
\affiliation{University of Chinese Academy of Sciences, Beijing 100049, China}

\author{Lun-hui Hu}
\affiliation{Center for Correlated Matter and School of Physics, Zhejiang University, Hangzhou 310058, China}

\author{Tiantian Zhang}
\affiliation{Institute of Theoretical Physics, Chinese Academy of Sciences, Beijing 100190, China}

\author{Jiangping Hu}
\affiliation{Beijing National Laboratory for Condensed Matter Physics and Institute of Physics, Chinese Academy of Sciences, Beijing 100190, China}
\affiliation{Kavli Institute for Theoretical Sciences and CAS Center for Excellence in Topological Quantum Computation, University of Chinese Academy of Sciences, Beijing 100190, China}
\affiliation{New Cornerstone Science Laboratory, Beijing, 100190, China}

\begin{abstract}
When the surface states of a topological insulator becomes superconducting, topological superconductivity can be obtained, and each vortex on the surface can host one single Majorana zero-energy mode which is usually a wave packet decaying exponentially off the vortex core. Here, we predict stable Majorana zero-energy mode whose wave function is ring-shape, dubbed as annular Majorana mode, in the superconducting vortex in topological insulators respecting $3$-fold or $6$-fold rotational symmetry. Such topological insulators are featured with a single nonlinear Dirac cone located at $\bar{\Gamma}$ or three linear Dirac cones at $\bar{\text{M}}$ in the surface Brillouin zone. The annular Majorana mode originates from the effective chiral $f$-wave superconductivity on the nonlinear Dirac cone in the former case and the interference of the effective chiral $p$-wave superconductivity on the three linear Dirac cones in the latter. In both cases, the annular Majorana mode is stabilized by the rotational symmetry and the winding number $3$ carried by the surface states. Candidate materials supporting the annular Majorana mode are predicted. Our work provides new insights into the topological superconductivity in superconducting topological insulators.
\end{abstract}


\maketitle


\textit{Introduction.}
Searching for new topological quantum states is one prominent goal in condensed matter physics. Topological superconductors\cite{RevModPhys.82.3045, RevModPhys.83.1057, RevModPhys.88.035005, Kitaev_2001, Alicea_2012} are intriguing, as they can host Majorana mode whose antimode is the mode itself. Due to the potential application in fault-tolerant quantum computations\cite{RevModPhys.80.1083, PhysRevX.5.041038, doi:10.1073/pnas.1810003115}, enormous endeavors have been devoted to pursuing the Majorana modes both theoretically\cite{PhysRevLett.100.096407, PhysRevLett.102.187001, PhysRevB.82.184516, PhysRevLett.105.097001, PhysRevLett.104.040502, PhysRevLett.105.077001, PhysRevLett.105.177002, PhysRevLett.105.046803, PhysRevB.82.115120, PhysRevLett.107.097001, PhysRevLett.111.047006, PhysRevLett.111.056402, PhysRevB.88.155420, PhysRevLett.115.127003, Li2016, PhysRevLett.117.047001, PhysRevB.93.224505, PhysRevX.9.011033, PhysRevLett.111.087002, doi:10.7566/JPSJ.82.113707, PhysRevLett.111.056403, PhysRevLett.115.187001, PhysRevLett.122.227001, PhysRevLett.119.246401, PhysRevB.97.205135, PhysRevLett.121.096803, PhysRevLett.121.186801, PhysRevLett.122.187001, PhysRevLett.122.126402, PhysRevX.10.041014, PhysRevX.12.011030, Zhang2024, PhysRevLett.133.106601, Zhang2026} and experimentally\cite{PhysRevLett.107.217001, Das2012, doi:10.1126/science.1259327, PhysRevLett.114.017001, PhysRevLett.116.257003, doi:10.1126/science.aao1797, Kong2019, Machida2019, chen2019observation, PhysRevX.8.041056, Kong2021, Li2022, Liu2024, doi:10.1126/science.aaw8419, doi:10.1126/science.aav3392} in the past decades. Superconducting vortex, topological defect in type-II superconductors, is an important platform in realizing the Majorana zero-energy mode (MZM). It has been long recognized MZMs can be bound in vortices in chiral $p$-wave superconductors\cite{PhysRevB.61.10267, PhysRevB.44.9667}. However, due to the rareness of the chiral superconductors, evidence for MZMs in chiral superconductors still lacks. The Fu-Kane formula provides an alternative route to the topological superconductors based on conventional superconductivity\cite{PhysRevLett.100.096407}. It is proposed that in proximity to conventional superconductors, the surface states of a topological insulator become topological superconducting and each vortex can bind one single MZM on the surface\cite{PhysRevLett.100.096407}. Subsequently, the investigation is extended to other systems\cite{PhysRevLett.112.106401, PhysRevLett.123.027003, PhysRevLett.122.207001, PhysRevLett.127.187002, 10.1093/nsr/nwac121, QIN20191207, PhysRevLett.124.257001, PhysRevLett.125.037001, PhysRevB.102.180505, PhysRevB.103.L140502, PhysRevLett.129.277001, PhysRevB.107.214518, Hu2023, Zhang2025}, and vortex bound states with exotic topological properties have been revealed in superconducting topological crystalline insulators\cite{PhysRevLett.112.106401} and topological semimetals\cite{PhysRevLett.123.027003, PhysRevLett.122.207001, PhysRevLett.127.187002, 10.1093/nsr/nwac121}. In experiments, signatures for vortex bound MZMs have been observed in superconductor heterostructure Bi$_2$Te$_3$/NbSe$_2$\cite{PhysRevLett.114.017001, PhysRevLett.116.257003}, FeTe$_{0.55}$Se$_{0.45}$\cite{doi:10.1126/science.aao1797, Kong2019, Machida2019, chen2019observation}, (Li$_{0.84}$Fe$_{0.16}$)OHFeSe\cite{PhysRevX.8.041056}, LiFeAs\cite{Kong2021, Li2022}, SnTe\cite{Liu2024}, etc.

\begin{figure}[!htbp]
	\centering
	\includegraphics[width=.98\linewidth]{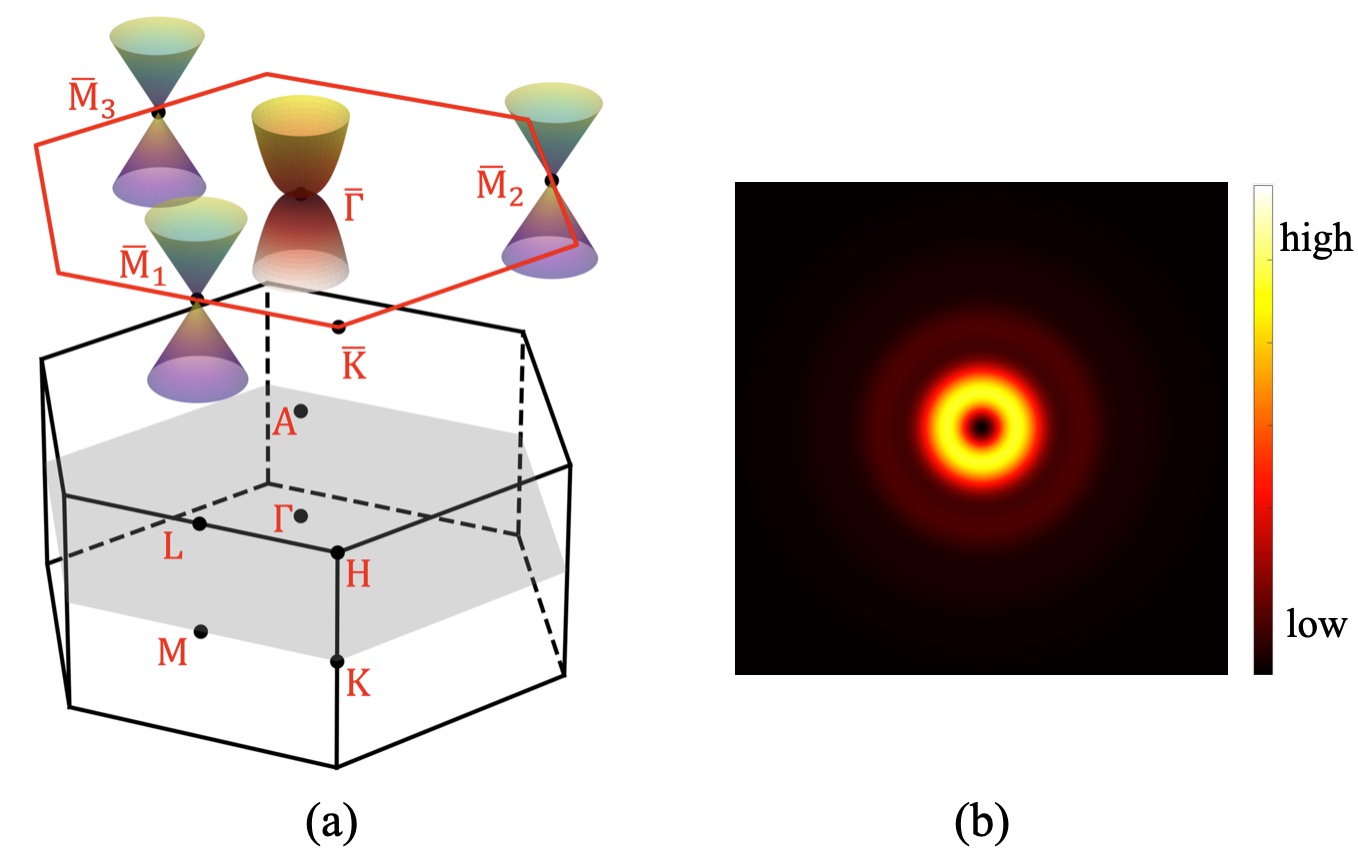}
	\caption{\label{fig1} (color online) Annular Majorana mode can be realized in the superconducting vortices in topological insulators with $3$-fold or $6$-fold rotational symmetry. (a) sketches the two conditions for the surface states supporting the annular Majorana mode, i.e. one single nonlinear Dirac cone located at the $\bar{\Gamma}$ point or three linear Dirac cones at the three $\bar{\text{M}}$ points in the $(001)$ surface BZ. (b) illustrates the real-space wave function profiles for the annular Majorana mode.}
\end{figure}

In current studies, especially in experimental detections\cite{PhysRevLett.114.017001, PhysRevLett.116.257003, doi:10.1126/science.aao1797, Kong2019, Machida2019, chen2019observation, PhysRevX.8.041056, Kong2021, Li2022, Liu2024}, the vortex bound MZM is always a wave packet decaying exponentially off the vortex core. Here, we predict a new type of MZMs which have a ring shape, dubbed as annular Majorana mode as illustrated in Fig.\ref{fig1}(b), existing in the superconducting vortices in $3$-fold or $6$-fold rotational symmetric topological insulators whose surface states carry a winding number $3$. Specifically, the topological surface states can be a single nonlinear Dirac cone located at $\bar{\Gamma}$ or three linear Dirac cones located at $\bar{\text{M}}$ as sketched in Fig.\ref{fig1}(a), which are both stabilized by the rotational symmetry and the time reversal symmetry in the surface Brillouin zone (BZ). Actually, the annular Majorana mode can be chirality-selected realized in vortices in chiral $p$-wave superconductors\cite{SuppMat}. Whereas, the annular Majorana mode is robust here. It stems from the effective chiral $f$-wave superconductivity on the nonlinear Dirac surface states, or the interference of the effective chiral $p$-wave superconductivity on the three Dirac cones on the surface.


\textit{Topological insulators with nonlinear surface states.}
We first focus on the $C_{6z}$ symmetric topological insulator, which has a single nonlinear Dirac cone at the BZ center on the surface. The surface states are protected by the time reversal symmetry $T$ and the $C_{6z}$ symmetry, and its basis is comprised of one Kramers' pair with angular momentum $J_z = \pm 3/2$ with $j_z$ the angular momentum defined according to $C_{6z}$. On the surface, the matrix form of the symmetry operations are $\widetilde{\mathcal{C}}_{6z} = e^{is_3\pi/2}$ and $\widetilde{\mathcal{T}} = is_2 K$ with $K$ being the complex conjugation operation. Based on a symmetry analysis, we can get the low-energy effective Hamiltonian for the surface states
\begin{align}\label{eq_surface}
h_{surf}({\bm k}) &= i \tilde{A} ( k_-^3 s_+ - k_+^3 s_- ) + i \tilde{B} ( k_+^3 s_+ - k_-^3 s_- ) + \tilde{M} k^2,
\end{align}
with $k_\pm = k_x \pm ik_y$, $k^2 = k_+k_-$ and $s_\pm = (s_1 \pm is_2)/2$. For the above surface states, one can calculate the winding number $w$ on the Fermi surface which turns out to be $3$ in accordance with the off-diagonal winding term, while the surface states disperse quadratically near the Dirac point due to the kinetic energy term. It is worth mentioning that the kinetic energy term in Eq.\eqref{eq_surface} does not destroy the surface anomaly because it has lower order compared to the winding term.

The topological insulator hosting the nonlinear surface Dirac cone described by $h_{surf}$ in Eq.\eqref{eq_surface} can be realized from band inversion between two Kramers' pairs $|J_z = \pm 3/2, \iota \rangle$ and $|J_z = \pm 3/2, -\iota \rangle$ at the BZ center in the bulk, where $\iota$ is the parity. The low-energy effective Hamiltonian for such a topological insulator takes the form
\begin{align}\label{eq_Hamiltonian}
h_{bulk}({\bm k}) &= m({\bm k})\sigma_3 + iA ( k_-^3 s_+\sigma_1 - k_+^3 s_-\sigma_1 ) \\ \nonumber
& + iB ( k_+^3 s_+\sigma_1 - k_-^3 s_-\sigma_1 ) + C k_z s_3\sigma_1,
\end{align}
where $m({\bm k}) = M_0 + M_1 k^2 + M_2 k^4 + M_3 k_z^2$, and $s$ and $\sigma$ are the Pauli matrices in the spaces spanned by $J_z$ and $\iota$ respectively. The Hamiltonian in Eq.\eqref{eq_Hamiltonian} is in the topological insulating state when the parameters satisfy $M_0 M_2 < 0$ and $M_0 M_3 < 0$. The corresponding matrix forms of the symmetry operators for the topological insulator are $\mathcal{C}_{6z} = e^{is_3\pi/2}$, $\mathcal{T} = is_2 K$ and the inversion symmetry $\mathcal{I} = \sigma_3$. Notice that in Eq.\eqref{eq_Hamiltonian} the $k^4$ term needs to be taken into consideration, to make sure the bulk Hamiltonian $h_{bulk}$ has no anomaly and converges as $k \rightarrow \infty$.

\textit{Annular Majorana mode from nonlinear surface Dirac cone.}
We consider conventional superconductivity in the nonlinear surface Dirac cone in Eq.\eqref{eq_surface}. The superconductivity can be intrinsic like the condition of the iron-based superconductors, or extrinsic from the proximity effect. Projecting the superconductivity onto the band basis, one can find that on the Fermi surface the superconductivity is equivalent to the chiral $f$-wave pairing in accordance with the winding number $3$ carried by the surface states\cite{SuppMat}. Consequently, topological superconductivity can be expected and MZMs can be realized in the vortices on the surface of the above topological insulator.

We simulate the vortex bound states straightforwardly, assuming conventional superconductivity in the surface states in Eq.\eqref{eq_surface}. The corresponding BdG Hamiltonian takes the form
\begin{align}\label{eq_surface_BdG}
\mathcal{H}_{sc}({\bm k}) &= [h_{surf}({\bm k}) - \mu] \tau_3 + \Delta({\bm r}) \tau_+ + \Delta^\dagger({\bm r}) \tau_-,
\end{align}
in the basis $\psi^\dagger({\bm k}) = ( d^\dagger ({{\bm k}}), is_2 d ({-{\bm k}}) )$ with $d^\dagger ({{\bm k}})$ being the basis for $h_{surf}({\bm k})$. In Eq.\eqref{eq_surface_BdG}, $\mu$ is the chemical potential, and $\tau$ is the Pauli matrix in the Nambu space with $\tau_\pm = (\tau_1 \pm i\tau_2)/2$. In the presence of a vortex, the superconducting order takes the form $\Delta({\bm r}) = \Delta_0 \tanh(r/
\xi) e^{i\theta}$ in the polar coordinates $(r, \theta)$, where $\Delta_0$ is the superconducting gap far from the vortex and $\xi$ characterizes the vortex size. 

We consider the symmetry constraints on the vortex bound states. we first look at a special case $\tilde{B} = 0$ in Eq.\eqref{eq_surface}. In this case, the surface states respect the continuum rotational symmetry rather than the $6$-fold rotation. As a result, the superconducting vortex bound states can be classified according to its the angular momentum $l_z$, which takes all integer value due to the flux of the vortex. Considering that the Majorana mode is particle-hole invariant and the particle-hole symmetry inverts the angular momentum $l_z$ to $-l_z$, the MZMs can only exist in the channel with $l_z = 0$. When $\tilde{B} \neq 0$, the vortex bound states is classified according to the $6$-fold rotation, which takes the eigenvalues $e^{il_z\pi/3}$. Namely, the states featured with different $l_z$ but the same $l_z$ $mod$ $6$ can couple with each other, and the MZMs must exist in the subspace with $l_z$ $mod$ $6 = 0$ or $3$. 


\begin{figure}[!htbp]
	\centering
	\includegraphics[width=.98\linewidth]{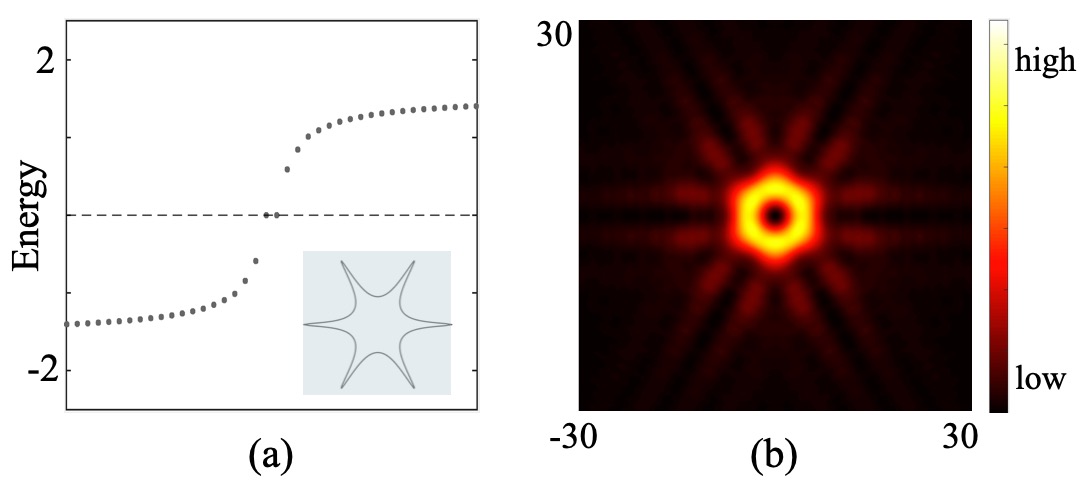}
	\caption{\label{fig2} (color online) (a) shows the low-energy superconducting vortex bound states in the subspace with $l_z$ $mod$ $6 = 0$, corresponding to the nonlinear surface Dirac cone depicted by $h_{surf}$ in Eq.\eqref{eq_surface}. The inset illustrates the Fermi surface utilized in the simulations. (b) presents the wave function profiles for the zero-energy mode in (a). In the calculations, the parameters are set to be $\tilde{M} = 0$, $\tilde{A} = 50$, $\tilde{B} = 45$, $\Delta_0 = 1.5$, $\mu = 2$ and $\xi = 10$.}
\end{figure}


Based on Eq.\eqref{eq_surface_BdG}, we solve the system numerically and get the vortex bound states in different $C_{6z}$ invariant subspaces. Similar to the case in the Fu-Kane formula, we find one single zero-energy mode living in the $l_z$ $mod$ $6 = 0$ channel in the vortex as presented in Fig.\ref{fig2}(a), which is undoubtedly the MZM; However, the MZM in our consideration has completely different real-space wave function profiles. As shown in Fig.\ref{fig2}(b), the MZM has a ring-like shape, different from the previous studies where the vortex bound MZM is a wave packet decaying exponentially off the vortex core.

The appearance of the annular Majorana mode is a direct result of the high winding number of the nonlinear surface states. To show this, it is instructive to consider the condition $\tilde{B} = 0$ where $\mathcal{H}_{sc}$ can be decoupled with respect to the angular momentum. It is convenient to express the eigenfuctions in the form of the Bessel functions in each subspace, and in the subspace with angular momentum $l_z = n$ the eigenfunctions read as
\begin{align}\label{eq_surface_VBS}
\phi_n({\bm r}) &= \sum_j ( e^{-i\theta} a_{1j} \tilde{J}_{n-1}(\beta^{n-1}_j \tilde{r}), e^{2i\theta} a_{2j} \tilde{J}_{n+2}(\beta^{n+2}_j \tilde{r}), \\ \nonumber
& \ e^{-2i\theta} a_{3j} \tilde{J}_{n-2}(\beta^{n-2}_j \tilde{r}), e^{i\theta} a_{4j} \tilde{J}_{n+1}(\beta^{n+1}_j \tilde{r}) )^T  e^{i n \theta} / \sqrt{2\pi},
\end{align}
where $\tilde{r} = r/R$ with $R$ the radius of the system, $\tilde{J}_{n}(\beta^{n}_j \tilde{r})$ is the $n$-th order normalized Bessel function of the first kind with $\beta^{n}_j$ its $j$-th zero, and $a_{ij}$ ($i = 1, 2, 3, 4$) is the corresponding coefficient. It is worth pointing out that the form of $\phi_n({\bm r})$ in Eq.\eqref{eq_surface_VBS} is closely related to the winding term in Eq.\eqref{eq_surface}. Recalling that the MZM exists in the $l_z = 0$ channel, one can find that the wave function profile of the MZM must vanish at the vortex core, i.e. $| \phi_0({\bm r = 0}) | = 0$, as $\tilde{J}_{n} (0) = 0$ for $n \neq 0$. Namely, the MZM from the nonlinear surface Dirac cone in Eq.\eqref{eq_surface} is a ring. When a finite $\tilde{B}$ term is tuned on, the wave function of the MZM has the following general form
\begin{align}\label{eq_surface_VBS_C6}
\phi_{MZM}({\bm r}) &= \sum_{(n \ mod \ 6) = 0} b_n \phi_n({\bm r}),
\end{align}
which shows the MZM maintains the ring-shape. In fact, the $\tilde{B}$ term merely introduces hexagonal distortions to the annular Majorana mode, as indicated in Fig.\ref{fig2}(b).


\textit{Annular Majorana mode from linear surface Dirac cone.}
Besides the above nonlinear surface states, an alternative condition which realizes the topological surface states carrying winding number $3$ is that at the three distinct $\bar{\text{M}}$ points in the surface BZ locate three linear Dirac cones which are related by the the $6$-fold or $3$-fold rotational symmetry. In the following, we shall show that due to the interference of the three Dirac cones, the annular Majorana mode exists in the superconducting vortex when conventional superconductivity is introduced into the surface. 

Prior to analyzing the superconducting vortex bound states, it is worth pointing out that there is no limiting conditions for the band inversion here. For simplicity, we still take the topological insulator in the previous condition, which has band inversion between Kramers' pairs carrying $J_z = \pm 3/2$ at the BZ center, as an concrete instance in the following analysis (other cases realizing such surface states are discussed in the SM\cite{SuppMat}). Generally, the $\mathcal{Z}_2$ index of the topological insulator merely demands Dirac cone at the time reversal invariant momentum in the surface BZ and has no further hard requirement on its specific location. The above topological insulator can support three Dirac cones at the $\bar{\text{M}}$ points.

Now, we consider the low-energy effective theory for the surface states. At each $\bar{\text{M}}$ point, the system respects the time reversal symmetry $\widetilde{\mathcal{T}} = is_2 K$ and the $2$-fold rotational symmetry $\widetilde{\mathcal{C}}_{2z} = is_3$. Taking these symmetry constraints into account, we achieve the general low-energy effective Hamiltonian for the Dirac cone at $\bar{\text{M}}_1$
\begin{align}\label{eq_surface_linear}
h_{1}({\bm k}) &= i \tilde{\mathcal{A}} ( k_- s_+ - k_+ s_- ) + i \tilde{\mathcal{B}} ( k_+ s_+ - k_- s_- ).
\end{align}
As the three Dirac cones are related by $C_{3z}$, namely $d_{2,3}(C_{3z} {\bm k}) = C_{3z} d_{1,2}({\bm k}) C_{3z}^{-1}$ with $d_{i}({\bm k})$ being the basis at $\bar{\text{M}}_i$ ($i = 1, 2, 3$),
the effective Hamiltonian for surface states at $\bar{\text{M}}_2$ and $\bar{\text{M}}_3$ can be obtained by applying $C_{3z}$ on $h_{1}({\bm k})$ in Eq.\eqref{eq_surface_linear}. Notice that for the effective theory $h_{i}({\bm k})$, ${\bm k}$ is measured from $\bar{\text{M}}_i$ respectively and $d_{i}({\bm k})$ corresponds to the slow-oscillating part of the wave function.

\begin{figure}[!htbp]
	\centering
	\includegraphics[width=.7\linewidth]{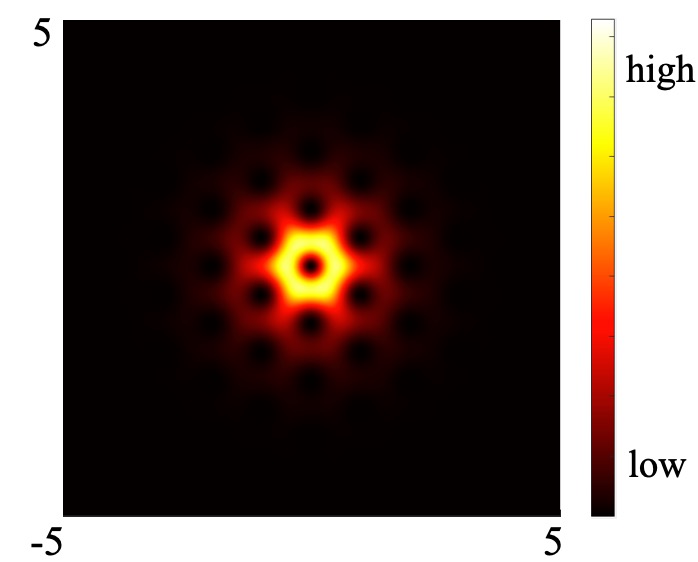}
	\caption{\label{fig_analyticM} (color online) The wave function profiles for the MZM  $| \gamma({\bm r}) |^2$ in the superconducting vortex in the condition with three linear surface Dirac cones at the $\bar{\text{M}}$ points. The figure is plotted based on the analytic result in Eq.\eqref{eq_WaveF_MZM_final}. In the numerical simulation, we take a simple case $\Delta(r) = \Delta_0$ and set $\Delta_0/\tilde{\mathcal{A}} = 0.8$.}
\end{figure}

When conventional superconductivity is introduced, on each Dirac cone the superconductivity is the effective chiral $p$-wave. Thus, in the superconducting vortex each Dirac cone would contribute one single MZM whose wave function decays exponentially away from the vortex core. When all the three Dirac cones are taken into account, the MZMs in a vortex will hybridize with two gapped out and one left stable. To unveil the feature of the MZM left in the vortex, we solve the system analytically. We first ignore the hybridization and treat the three Dirac cones separately. Actually, the analytical wave function for the MZM originating from the Dirac cone described by Eq.\eqref{eq_surface_linear} is hard to obtain. However, we can focus on the specific situation with $\tilde{\mathcal{B}} = 0$ as the system can always be tuned to such a condition adiabatically in the superconducting state. Assuming ${\tilde{\mathcal{A}}} \Delta_0 > 0$, the wave functions of the MZMs take a simple form at the chemical potential $\mu = 0$ as follows (more detailed analysis in the SM\cite{SuppMat})
\begin{align}\label{eq_WaveF_MZM}
\gamma_1 &= \int d^2{\bm r} [ d_{1,\downarrow}({\bm r}) + d_{1,\downarrow}^\dagger({\bm r}) ] e^{ -\int_0^r \frac{\Delta(r^\prime)}{\tilde{\mathcal{A}}} dr^\prime }, \\ \nonumber
\gamma_2 &= \int d^2{\bm r} [ e^{2i\pi/3} d_{2,\downarrow}({\bm r}) + e^{-2i\pi/3} d_{2,\downarrow}^\dagger({\bm r}) ] e^{ -\int_0^r \frac{\Delta(r^\prime)}{\tilde{\mathcal{A}}} dr^\prime }, \\ \nonumber
\gamma_3 &= \int d^2{\bm r} [ e^{4i\pi/3} d_{3,\downarrow}({\bm r}) + e^{-4i\pi/3} d_{3,\downarrow}^\dagger({\bm r}) ] e^{ -\int_0^r \frac{\Delta(r^\prime)}{\tilde{\mathcal{A}}} dr^\prime },
\end{align}
where $\gamma_i$ is the MZM resulting from the Dirac cone at $\bar{\text{M}}_i$ and the normalization factor is omitted for simplicity. Note that in Eq.\eqref{eq_WaveF_MZM} the phases in $\gamma_i$ are contributed by two parts, including the rotation of the vortex and the rotation of the spinor ($J_z = \pm 3/2$) with respect to $\bar{\Gamma}$.



Then, we consider the hybridization among the three MZMs, by taking the symmetry constraints into account. Under the $3$-fold rotation, the MZMs in Eq.\eqref{eq_WaveF_MZM} transforms as $\gamma_1 \rightarrow \gamma_2$, $\gamma_2 \rightarrow \gamma_3$, $\gamma_3 \rightarrow \gamma_1$. Accordingly, the rotational-symmetric hybridization can be written as 
\begin{align}\label{eq_MZM_hybridize}
h_{hyb} = i\eta ( \gamma_1\gamma_2 + \gamma_2\gamma_3 + \gamma_3\gamma_1 ).
\end{align}
Straightforwardly, we solve $h_{hyb}$ and the MZM left in the vortex turns out to be $\gamma = \gamma_1 + \gamma_2 + \gamma_3$. Recalling that the Dirac cones are located at the $\bar{\text{M}}$ points, we get the real-space wave function for the MZM\cite{footnote1}
\begin{align}\label{eq_WaveF_MZM_final}
\gamma({\bm r}) &= e^{ i{\bm K}_{1} \cdot {\bm r} } \gamma_1({\bm r}) + e^{ i{\bm K}_{2} \cdot {\bm r} } \gamma_2({\bm r}) + e^{ i{\bm K}_{3} \cdot {\bm r} } \gamma_3({\bm r}),
\end{align}
with ${\bm K}_{i}$ the momentum corresponding to $\bar{\text{M}}_i$. From Eqs.\eqref{eq_WaveF_MZM}\eqref{eq_WaveF_MZM_final}, it can be found $| \gamma({\bm r}) | = 0$ due to the coherent destructive interference effect of the three Dirac cones near the vortex core, indicating the ring shape of the MZM in the vortex. We plot the analytical wave function in Eq.\eqref{eq_WaveF_MZM_final} in Fig.\ref{fig_analyticM}, which shows the MZM does has the ring shape.

\begin{figure}[!htbp]
	\centering
	\includegraphics[width=.999\linewidth]{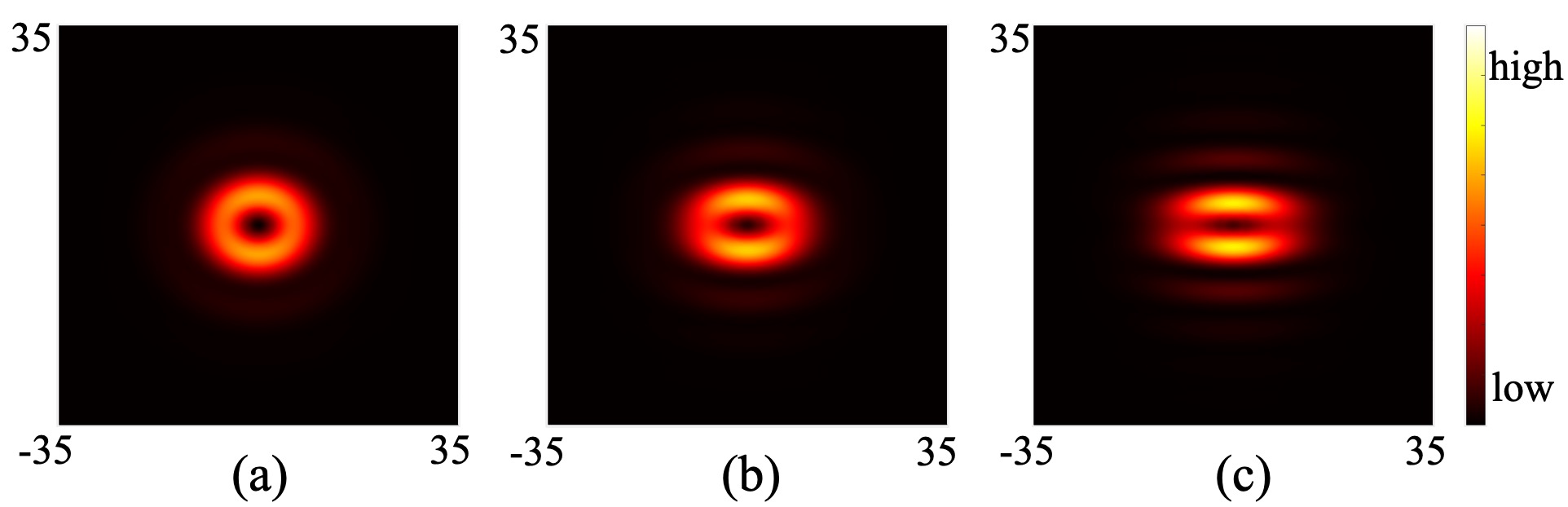}
	\caption{\label{fig3} (color online) (a)$\sim$(c) show the wave function profiles for the annular Majorana mode for the nonlinear Dirac surface states $h_{surf}$ in Eq.\eqref{eq_surface} in the presence of the nematic perturbation $h_{nem}$ in Eq.\eqref{eq_nematic}. The figures share the same color bar. In the calculations, we set the nematic perturbations to be $\tilde{D} = 2$ in (a), $\tilde{D} = 5$ in (b) and $\tilde{D} = 8$ in (c). The other parameters are chosen as $\tilde{M} = 0$, $\tilde{A} = 50$, $\Delta_0 = 1.5$, $\mu = 2$ and $\xi = 10$. }
\end{figure}

\textit{Symmetry breaking and multi-lobe Majorana mode.}
We discuss the effects of the symmetry-breaking perturbations on the annular Majorana mode. As the rotational symmetry plays an essential role in stablizing the annular Majorana mode, we focus on nematic perturbation which breaks the rotational symmetry from $C_{6z}$ to $C_{2z}$ and maintains the time reversal symmetry. Such perturbations may arise from external uniaxial pressure or spontaneous symmetry breaking. We utilize the nonlinear surface states in Eq.\eqref{eq_surface} to carry out numerical simulations. Based on a symmetry analysis, the leading-order nematic perturbation in the surface states is obtained
\begin{align}\label{eq_nematic}
h_{nem}({\bm k}) &= \tilde{D} ( k_- s_+ + k_+ s_- ).
\end{align}
In the presence of $h_{nem}({\bm k})$, the nonlinear surface Dirac cone is no longer stable, and it will split into three linear ones with the total winding number on the Fermi surfaces preserved. Based on $h_{surf} + h_{nem}$, we simulate the superconducting vortex bound states, and in the calculations we set $\tilde{B} = \tilde{M} = 0$ for simplicity. The numerical results are presented in Figs.\ref{fig3}(b)$\sim$(d). As shown, the annular Majorana mode becomes more and more anisotropic as the nematicity is stronger, and the ring-shape wavefunction is torn up forming discrete lobes in real space when the perturbation is strong enough. This is different from the conventional case, where such perturbations merely introduce some anisotropy into the MZM wave packet.

 \begin{figure}[!htbp]
 	\centering
 	\includegraphics[width=.9\linewidth]{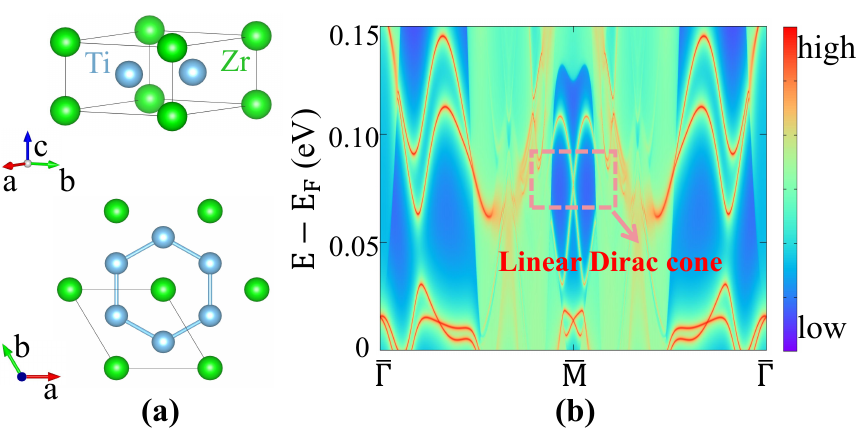}
 	\caption{\label{fig_ZrTi2} (color online) (a) depicts the crystal structure of ZrTi$_2$. (b) shows the surface states of ZrTi$_2$ on the $(001)$ surface, where the surface is terminated at the layer composed of the Ti atoms, showing a linear Dirac cone located at each $\bar{\text{M}}$ point.}
 \end{figure}

\textit{Candidate materials.}
We predict ZrTi$_2$ a topological insulator whose surface states carrying winding number $3$. ZrTi$_2$ is a nonmagnetic three-dimensional material that crystallizes in space group P6/mmm (No.191). It has a layered crystal structure, which consists of layers of Zr and layers Ti. Near the Fermi level, the energy bands are mainly contributed by the $d$ orbitals of Zr and Ti. As the system is inversion-symmetric, its topological property can be conveniently identified by the parities of the occupied bands at the time-reversal invariant momenta. Based on first-principle calculations, we find ZrTi$_2$ has a band inversion between the $\Gamma^{-}_{10}$ and $\Gamma^{+}_{7}$ bands at $\Gamma$\cite{PhysRevB.74.195312,bradlyn2017topological,zhang2019catalogue}. The $\Gamma^{-}_{10}$ and $\Gamma^{+}_{7}$ bands are Kramers' doublets both carrying angular momenta $J_z = \pm 3/2$ but the opposite parities. The band inversion leads to the topological invariant $Z_2 = 1$ in ZrTi$_2$. Moreover, straightforward simulations show that the surface states of ZrTi$_2$ are composed of three linear Dirac cones at the three $\bar{\text{M}}$ points in the surface BZ, which locate about $50$ meV above the Fermi level on the surface terminated at the Ti layer, as shown in Fig.\ref{fig_ZrTi2}. On the surface terminated at the Zr layer, the Dirac cones are at $\bar{\text{M}}$ and are about $80$ meV above the Fermi level. We present more detailed calculations on ZrTi$_2$ in the SM\cite{SuppMat}. In accordance with our theory, ZrTi$_2$ can host the annular Majorana mode in the superconducting vortex.

\textit{Discussion and summary.}
Until now, we have focused on topological insulators with $6$-fold rotational symmetry. In fact, the above surface states and annular Majorana modes can also be supported in topological insulators respecting the $3$-fold rotational symmetry. As the linear surface Dirac cones at $\bar{\text{M}}$ is natural, we emphasize the existence of the nonlinear surface Dirac cones in $C_{3z}$ symmetric topological insulators. On the surface of such a topological insulator, $J_z = 3/2$ is actually equivalent to $J_z = -3/2$ according to $C_{3z}$; However, the time reversal symmetry guarantees the orthogonality of the two states, helping stabilize the nonlinear surface Dirac cone.

Besides, our theory can be extended to two-dimensional spin-$1/2$ and spin-$3/2$ Rashba electron gases which respect the $6$-fold or $3$-fold rotational symmetry. An external Zeeman field can destroy the Kramers' degeneracy in the electron gas. When conventional superconductivity is introduced, the annular Majorana mode is supported in the superconducting vortex if there are three nondegenerate Fermi surfaces at the three ${\text M}$ points in the spin-$1/2$ electron gas; And in the spin-$3/2$ system, the annular Majorana mode is supported if it has three nondegenerate Fermi surfaces at the three ${\text M}$ points or one nondegenerate Fermi surface at the $\Gamma$ point. We verify the scenario numerically in the SM\cite{SuppMat}.


Experimentally, a peak conductance off the vortex core combined with vanishing conductance at the vortex core at zero bias can provide strong evidence for the annular Majorana mode in scanning tunneling microscope measurements. However, we point out that in the case with three linear Dirac cones at $\bar{\text{M}}$, the energy spacing between the annular Majorana mode and other states can be rather small. This is due to the fact that the hybridization in Eq.\eqref{eq_MZM_hybridize} arises from the scattering between different $\bar{\text{M}}$ points, and such large momentum scattering leads to small hybridization coefficient $\eta$ as the vortex size is usually much larger than the lattice constant. Therefore, the other two quasiparticles in Eq.\eqref{eq_MZM_hybridize} can be nearly degenerate with the annular Majorana mode; And to distinguish these modes, ultra-high resolution measurements at ultra-low temperature are necessary in this condition.

In summary, we predict stable annular Majorana mode in superconducting vortex in the topological insulators with $3$-fold or $6$-fold rotational symmetry. Such kind of topological insulators are featured with surface states carrying total winding number $3$, which can be realized by a single nonlinear Dirac cone at $\bar{\Gamma}$ or three linear Dirac cones at $\bar{\text{M}}$ in the surface BZ. We predict ZrTi$_2$ a candidate material realizing the vortex bound annular Majorana mode. Our work provides new insights into the topological superconductivity in superconducting topological insulators.



\textit{Acknowledgment.} This work is supported by the National Natural Science Foundation of China (Grant No. NSFC-12304163, No. NSFC-11888101, No. NSFC-12174428, No. NSFC-11920101005), the Ministry of Science and Technology (Grant No. 2022YFA1403900), the Strategic Priority Research Program of the Chinese Academy of Sciences (Grant No. XDB28000000, XDB33000000, XDB1720000), the New Cornerstone Investigator Program, the Beijing Institute of Technology Research Fund Program for Young Scholars, the Chinese Academy of Sciences Project for Young Scientists in Basic Research (2022YSBR-048), and  National Key R\&D Project (Grant Nos. 2023YFA1407400 and 2024YFA1409200). L.-H. Hu is supported by National Key R\&D Program of China (Grant No. 2025YFA1411501), the National Natural Science Foundation of China (Grant No. NSFC-12561160109, No. NSFC-12574148).

\textit{Note added.} We notice that the nonlinear surface Dirac cones in topological insulators are systematically classified in a recent work\cite{shi2026symmetry}, but the topological superconductivity in such topological insulators is not involved in the work.

\nocite{*}
\bibliography{references}

\end{document}